\documentclass[5p]{elsarticle}
 
\usepackage{subfigure}
\usepackage{url}
\usepackage{graphicx}
\usepackage{float}
\usepackage{array}
\usepackage{textcomp}

\begin{document}

\begin{frontmatter}

\title{CMB-GAN: Fast Simulations of Cosmic Microwave background Anisotropy maps using Deep Learning}

\author[1]{Amit Mishra} 
\author[1]{Pranath Reddy}
\author[1]{Rahul Nigam}

\address[1]{Birla Institute of Technology \& Science Pilani - Hyderabad Campus, Hyderabad, India}

\begin{abstract}
Cosmic Microwave Background (CMB) has been a cornerstone in many cosmology experiments and studies since it was discovered back in 1964. Traditional computational models like CAMB that are used for generating CMB temperature anisotropy maps are extremely resource intensive and act as a bottleneck in cosmology experiments that require a large amount of CMB data for analysis. In this paper, we present a new approach to the generation of CMB temperature maps using a specific class of neural networks called Generative Adversarial Network (GAN). We train our deep generative model to learn the complex distribution of CMB maps and efficiently generate new sets of CMB data in the form of 2D patches of anisotropy maps without losing much accuracy. We limit our experiment to the generation of 56$^{\circ}$ and 112$^{\circ}$ square patches of CMB maps. We have also trained a Multilayer perceptron model for estimation of baryon density from a CMB map, we will be using this model for the performance evaluation of our generative model using diagnostic measures like Histogram of pixel intensities, the standard deviation of pixel intensity distribution, Power Spectrum, Cross power spectrum, Correlation matrix of the power spectrum and Peak count. We show that the GAN model is able to efficiently generate CMB samples of multiple sizes and is sensitive to the cosmological parameters corresponding to the underlying distribution of the data. The primiary advantage of this method is the exponential reduction in the computational time needed to generate the CMB data, the GAN model is able to generate the samples within seconds as opposed to hours required by the CAMB package with an acceptable value to error and loss of information. We hope that future iterations of this methodology will replace traditional statistical methods of CMB data generation and help in large scale cosmological experiments.
\end{abstract}


\end{frontmatter}

\label{sec:intro}
\section{Introduction}
The variations in temperature of the Cosmic Microwave Background (CMB) are similar to the ripples on the cosmic pond and enclose a lot of information about the universe. To collect this information we look at the scales at which these temperature fluctuations occur. The amount of temperature fluctuations (in micro Kelvin) is plotted against the multipole moment (l). This is the angular power spectrum graph of a CMB temperature map. Such graphs contain several peaks which provide a lot of information and we exploit this for our use. 

The first peak is an indication of the geometry of the universe, whether it is flat or curved (Hu, Wayne, et al., 2004).. CMB radiation is distorted by the curvature of the universe since the radiation comes from all directions of the visible universe. The fluctuations will appear undistorted if the universe is flat. The fluctuations would appear magnified if the universe is positively curved and de-magnified if it is negatively curved. The second peak reveals information about the number of baryons present in the universe. Due to the initial fluctuations in the universe, all matter would tend to gravitationally group towards the higher density fluctuations. However, baryon matter which is interactive with light would heat up as it clumps up, and the resultant pressure would try to push against the grouped matter. This implies that the second peak will be more damped if there is more matter. Thus, the ratio of the first and second peak gives us the baryon density(Bucher, M., 2015).

The anisotropy of the cosmic microwave background (CMB) consists of the small temperature fluctuations in the blackbody radiation left over from the Big Bang. The CMB temperature maps are an incredible source of information for cosmological analysis and the advent of big data methods (Alex Krizhevsky, Geoffrey E Hinton, 2012) have opened a new avenue for the analysis of CMB. Modern data analysis methods such as machine learning and deep learning require a large amount of data and traditional methods such as \textit{CAMB} and \textit{healpy} (Gorski, K. M., Hivon, E., Banday, A. J., et al. 2005) are computationally expensive and inefficient for generating a large number of CMB maps. Here we demonstrate the use of deep generative models to generate synthetic samples of CMB all-sky maps which can be used for cosmological analysis. Deep generative models are capable of learning complex distributions from a given dataset and then generate new, statistically consistent data samples (I. J. Goodfellow, 2014). We generate the dataset for the training of our generative model by snipping 128x128 and 256x256 resolution patches from the CMB maps which effectively gave us 56$^{\circ}$ and 112$^{\circ}$ patches respectively. We also train a multilayer perceptron network (MLP) to predict the baryon density of a given CMB map, this helps us in comparing the samples generated by the generative model with the samples of our dataset by correlating the baryon density predictions given by the MLP model. We use various diagnostic metrics like the histogram of pixel intensities, the standard deviation of pixel intensity distribution, Power Spectrum, Cross power spectrum, Correlation matrix of the power spectrum and Peak count to evaluate the performance of our generative model. The practical advantage of this method is that once the model has been trained, the generation process is extremely fast, thus giving us the ability to generate a large number of samples that can be used for scientific study. 

\section{Methodology}
\subsection{CAMB and Data Generation}
We use standard cosmological software \textit{CAMB} to generate CMB temperature maps for training. CAMB is used to compute CMB, CMB lensing and other related cosmological functions. CAMB takes several parameters as input to generate a file containing the initial angular power spectrum data of the universe. The Curved correlation function is used as the lensing method and we include reionization. Other physical parameters which are input to CAMB include Hubble constant, the temperature of CMB, baryon density, cold dark matter density, the effective mass density of dark energy, maximum multipoles data, redshift and helium fraction. This power spectrum file is in turn used by \textit{healpy} to generate random gaussian CMB temperature maps which are used for training the neural network. 

Anisotropy from dipole effect due to the movement of the earth relative to the CMB rest frame and galactic contaminants along the equator corresponding to the galactic plane is removed while generating the temperature maps. The generated full-sky maps have the galactic center at the center of the mollweide projection.

\begin{figure}[h]
	 
	\centering
	\includegraphics[width = 2.5in]{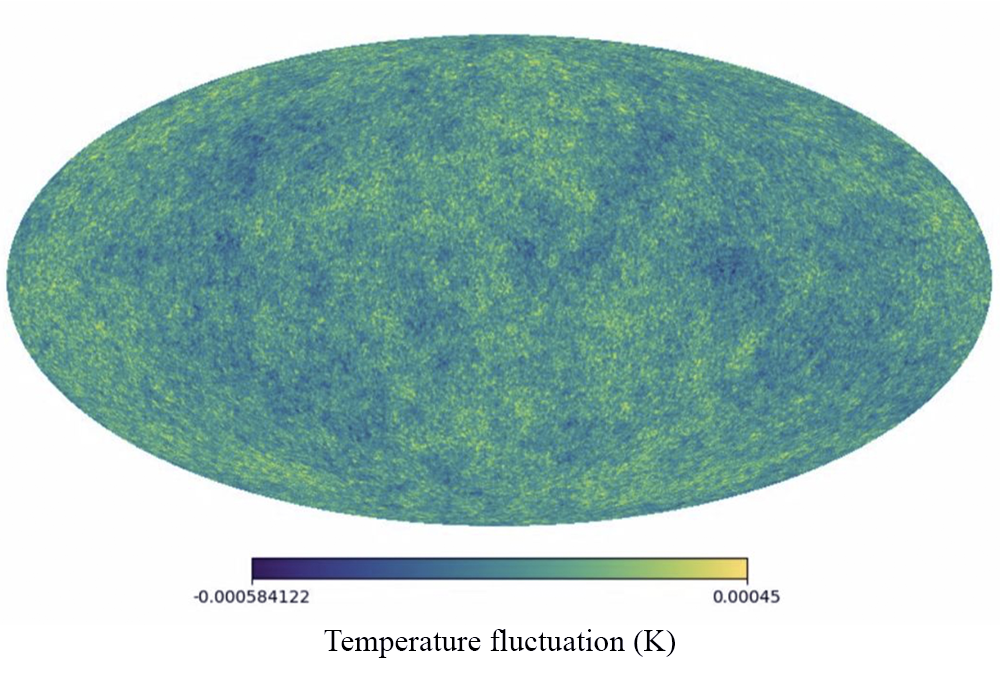}
	\caption{A random full sky CMB temperature map generated using healpy and CAMB}		 	
\end{figure}
\begin{figure}[!h]
	 
	\centering
	\includegraphics[width = 2.5in]{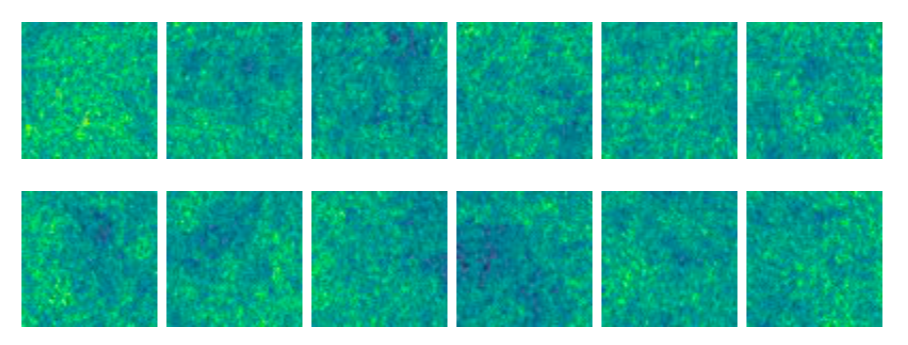}
	\caption{Sample patches cropped along the equator of the full sky CMB temperature map}		 	
\end{figure}

\subsection{Implementation and Training}
The method proposed in this paper comprises two steps, Baryon density estimation and CMB data generation using a traditional Artificial Neural Network and a Deep generative model. We use a Generative Adversarial Network trained on CMB patches obtained using CAMB for the generation of new CMB data and a Multilayer perceptron network trained on labeled CMB data for baryon density estimation which will be used for diagnosis and performance evaluation of our Generative network. We first train an image classifier using a multilayer perceptron network with the baryon density obtained from the power spectrum of CMB as the classes/labels of our data. Here we use a dataset with a large number of classes to approximate our classifier as a regression model, this helps us in predicting the baryon density of the input test images with a higher degree of precision. Convolutional neural networks (CNN) are one of the most famous set of neural network architectures used for classifying images, CNN takes advantage of local spatial coherence of the input (Rippel, Snoek, Adams, 2015) because we assume that the spatially close images used for training are correlated, but in the case of the CMB dataset , the pixels in the images are random noise following a gaussian distribution, the CNN network will not be able to find any common features in the inputs and thus the training accuracy and test error will be less than favorable. We have tested Resnet-101 and Inception v2 CNN architectures. The training accuracy of resnet network was very low whereas the inception network was suffering from high variance problem (Liu, Wei, Zhang, Yang, 2017). For this reason, we will be using a Multi-Layer perceptron network. A multilayer perceptron is one of the most commonly used architectures of feedforward artificial neural networks, it consists of three classes of layers and nodes, the input layer, hidden layers, and an output layer. Each node in a layer is connected to the nodes of the next layer via a non-linear activation function. Multilayer perceptron makes use of one of the most famous techniques of supervised learning called backpropagation (Goodfellow, Bengio, Courville, 2018) for training the network. A multilayer perceptron can be distinguished from a linear perceptron from its characteristic use of fully connected multilayers. This makes multilayer perceptrons suitable for working with non-linearly separable data (Bullinaria, 2015) and can be perceived as a logistic regression classifier. The weights of the fully connected layers are updated once a batch of data has been passed through the network by measuring the error of the output with the expected result (predetermined labels), this is the essence of learning in neural networks and is carried out with the help of an iterative algorithm called backpropagation. This is an example of supervised learning. Backpropagation uses an iterative optimization algorithm called gradient descent (Goodfellow, Bengio, Courville, 2018) to update the weights of the network. We continue to train the network until the training accuracy and the testing cost gets saturated. We have used a softmax cross-entropy function as our loss function. consider a mapping of input x to category y, we have\\

Objective: \[ min[-E_{x,y~p(data)}log(P(Y|X)] \] 
where, 	
\textit{E} is the expectation function\\
\textit{P(data)} is the true data distribution\\
\textit{P(y\textbar x)} is the distribution of our parametric model.\\ \\
We now train our generative model to generate the CMB data. The primary difference between a discriminative algorithm and a generative algorithm is that a discriminative algorithms map features to labels whereas a generative algorithm tries to predict the features given a certain label. Discriminative models learn the boundary between classes and Generative models model the distribution of individual classes. In this experiment, we use a Deep Convolutional Generative Adversarial Network which has the ability to mimic complex distributions of data. The primary goal of the Generative Adversarial Network is to generate new samples from the same distribution as that of the training data. The most notable feature of GAN is that it consists of a pair of networks: a generative network (G) and a discriminative network (D). The two networks are in a two-player game setting where the Generator network tries to fool the discriminator by generating images that match very closely to the training data and the Discriminator network tries to differentiate between real and generated images thus training jointly in a minimax game. The Discriminator tries to classify a sample x and outputs the likelihood in (0,1) of the real image, whereas the Generator uses a random variable z drawn from a given prior distribution.\\

Objective :
\small
		 \[min_{G}max_{D}[ E_{x \sim p(x)} \{log(D(x)\} + E_{z \sim  p(z)} \{log(1-D(G(z))\} ]\]	 	 						 			
where, \\ 	
E is the expectation function \\
P(x) is the true data distribution \\
P(z) is the prior distribution ( usually a Gaussian ) \\ 

The Discriminator D tries to maximize the objective such that D(\textit{x}) is close to 1 (real) and D(\textit{G(z)}) is close to 0 (fake) and the Generator G tries to minimize the objective such that D(\textit{G(z)}) is close to 1 (discriminator is fooled into thinking generated \textit{G(z)} is real). This training process is essentially trying to reduce the Jensen-Shannon divergence between \textit{P(x)} and \textit{P(z)}.
We have used the Tensorflow library to implement the MLP model and the GAN model. We have used Adam optimization (Kingma, Ba, 2017) algorithm instead of the traditional stochastic gradient descent for updating the weights of the network (Michelucci, Umberto, 2018) and used L2 regularization, also known as ridge regularization to prevent our model from overfitting. In L2 regularization, we add a squared error term as a penalty to the loss function (Goodfellow, Bengio, Courville, 2018). The training of the network is done in the Google Cloud platform using a Tesla K80 GPU.

\subsection{Network Configuration}
\subsubsection{MLP Network:}
\vspace{-1.5em}
\begin{table}[!h]
\small
\caption{Network Configuration}
\begin{tabular}{| >{\centering\arraybackslash}m{0.8in} | >{\centering\arraybackslash}m{1.5in} | >{\centering\arraybackslash}m{0.7in}|}
\hline
No of hidden layers & No of Nodes in each hidden layer & Learning Rate
\\
\hline
5 & 3223 & 0.001\\
\hline
\end{tabular}
\end{table}

\begin{table}[!h]
\small
\begin{tabular}{| >{\centering\arraybackslash}m{0.8in} | >{\centering\arraybackslash}m{1.3in} | >{\centering\arraybackslash}m{0.9in} |}
\hline
Batch size & No of epochs & Regularization parameter
\\
\hline
512  & 50000 & 0.01\\
\hline
\end{tabular}
\end{table}

The learning rate determines how fast the weights or the coefficients of the network are updated. An epoch can be defined as the number of times the algorithm perceives the entire data-set. Hence, an epoch is completed when all the samples of the data have been perused. An iteration can be defined as the number of times a “batch of data” has been passed through the algorithm. In the case of a multilayer perceptron, that means the forward pass and backward pass. Hence, an iteration is completed once a batch of data has passed through the network. The batch size is the number of training examples passed through the network at once (Shen, 2017 \& Svozil, Kvasnicka, Pospichal, 1997).

\subsubsection{GAN Network:}
We use a modified version (Alec Radford, 2015) of the standard GAN architecture incorporating convolution layer with a kernel Size of 5x5.

\begin{table}[!h]
\small
\caption{Discriminator Network Configuration}
\begin{tabular}{| >{\centering\arraybackslash}m{0.5in}|>{\centering\arraybackslash}m{0.85in}|>{\centering\arraybackslash}m{0.75in}|>{\centering\arraybackslash}m{0.5in} |}
\hline
No of hidden layers & Operations & Outputs & Batch Size
\\
\hline
5  (for 56$^{\circ}$ patches) 6  (for 112$^{\circ}$ patches) & Conv/ linear & Leaky-Relu-BatchNorm/ sigmoid & 50\\
\hline
\end{tabular}
\end{table}

\begin{table}[!h]
\small
\begin{tabular}{| >{\centering\arraybackslash}m{0.6in} | >{\centering\arraybackslash}m{0.8in} | >{\centering\arraybackslash}m{1.37in} |}
\hline
No of epochs & Learning Rate & Regularization parameter
\\
\hline
2000  & 0.000001 & 0.01\\
\hline
\end{tabular}
\end{table}

Dimension of the gaussian prior distribution ( linear input of generator) = 200

\begin{table}[!h]
\small
\caption{Generator Network Configuration}
\begin{tabular}{| >{\centering\arraybackslash}m{0.5in}|>{\centering\arraybackslash}m{0.85in}|>{\centering\arraybackslash}m{0.75in}|>{\centering\arraybackslash}m{0.5in} |}
\hline
No of hidden layers & Operations & Outputs & Batch Size
\\
\hline
5  (for 56$^{\circ}$ patches) 6  (for 112$^{\circ}$ patches) & linear/DeConv & Relu-BatchNorm/ tanh & 50\\
\hline
\end{tabular}
\end{table}

\begin{table}[!h]
\small
\begin{tabular}{| >{\centering\arraybackslash}m{0.6in} | >{\centering\arraybackslash}m{0.8in} | >{\centering\arraybackslash}m{1.37in} |}
\hline
No of epochs & Learning Rate & Regularization parameter
\\
\hline
2000  & 0.000001 & 0.01\\
\hline
\end{tabular}
\end{table}

\section{Results}
Here we present the results obtained after training the Multilayer Perceptron and the Generative adversarial network. We have focused our study on two classes of CMB simulations, 56$^{\circ}$ patches and 112$^{\circ}$ patches. We have generated 100 random samples using the trained GAN model and then passed them as the input to the trained MLP model to predict the corresponding baryon densities. We have then extracted the patches from CAMB training dataset whose baryon densities matches the predictions obtained from the MLP model and used them for comparison and evaluation in various metrics. The similarities in the power spectrum and other metrics have shown that out GAN model is capable of reconstructing the complex distribution underlying the CMB and the subtle but consistent discrepancies in the results ensure the model isn’t just memorizing the input data but rather recreating the distribution from the ground up. 
\subsection{56$^{\circ}$ patches}

\begin{figure}[!h]
	\centering
	\includegraphics[width = 3.2in]{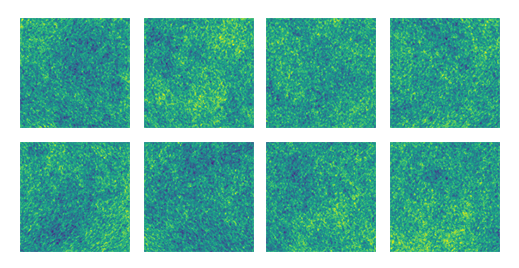}
	\caption{Sample patches (56$^{\circ}$s) from the maps generated using CAMB}		 	
\end{figure}

\begin{figure}[!h]
	\centering
	\includegraphics[width = 3.2in]{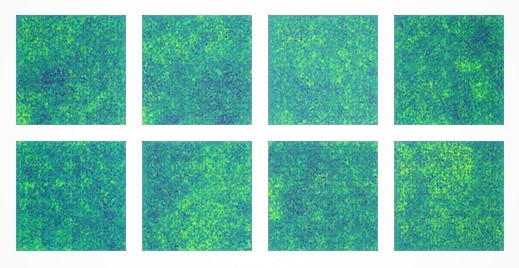}
	\caption{Sample patches (56$^{\circ}$s) generated by trained GAN model}		 	
\end{figure}

\begin{figure}[!h]
  \centering
  \begin{tabular}{@{}c@{}}
    \includegraphics[width = 3.2in]{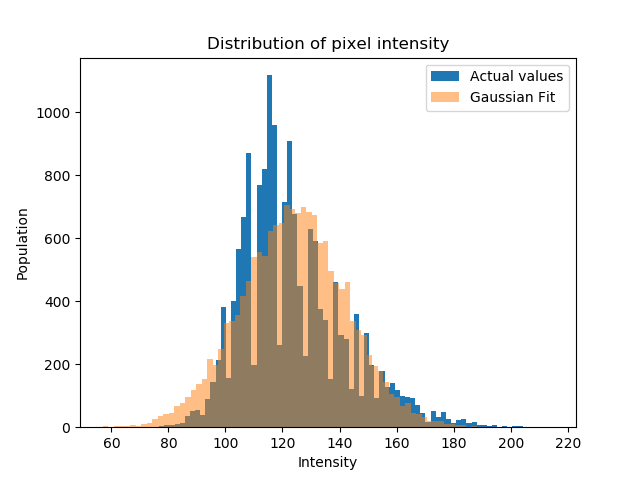} \\[\abovecaptionskip]
    \small (a)
  \end{tabular}

  \vspace{\floatsep}

  \begin{tabular}{@{}c@{}}
    \includegraphics[width = 3.2in]{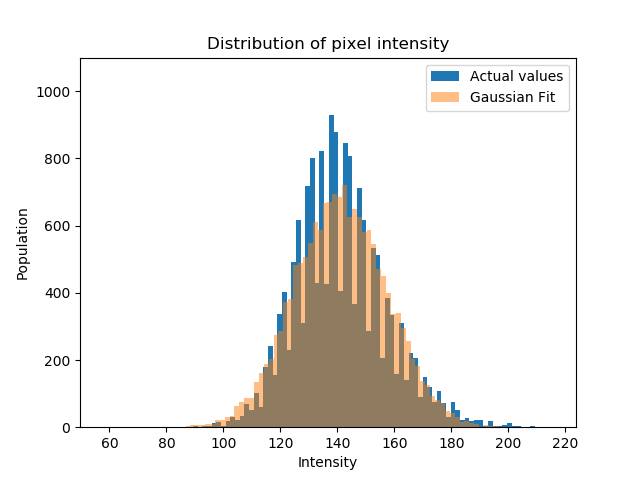} \\[\abovecaptionskip]
    \small (b) 
  \end{tabular}

  \caption{Distribution of pixel intensities of (a) a random sample patch from a map generated using CAMB and (b) a random sample patch generated using trained GAN model.}
\end{figure}

\newpage
Figures 5-11 represent the diagnostic results obtained from the 56$^{\circ}$ patches generated by the GAN model. \\
We have generated 100 random samples using the trained GAN model and then passed them as the input to the trained MLP model to predict the corresponding baryon densities. We have then extracted the patches from CAMB training dataset whose baryon densities matches the predictions obtained from the MLP model and used them for comparison and evaluation in various metrics. The similarities in the power spectrum and other metrics have shown that out GAN model is capable of reconstructing the complex distribution underlying the CMB and the subtle but consistent discrepancies in the results ensure the model isn’t just memorizing the input data but rather recreating the distribution from the ground up. 

\begin{figure}[!h]
	\centering
	\includegraphics[width = 3.5in]{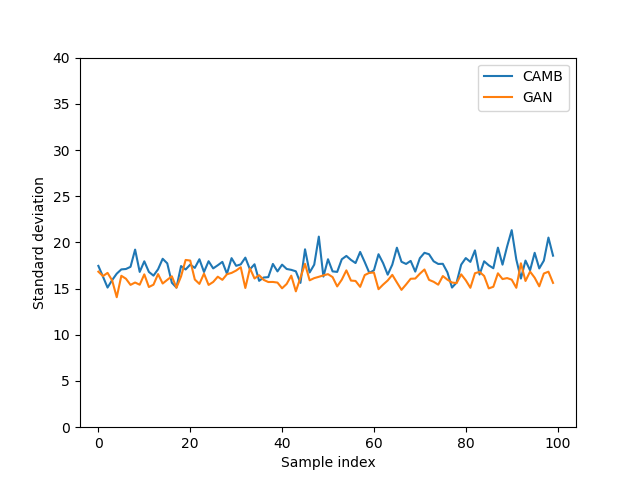}
	\caption{The plot of the standard deviation of the pixel intensity distribution of patches generated using GAN and the corresponding matched CAMB patches }		 	
\end{figure}

\begin{figure}[!h]
	\centering
	\includegraphics[width = 3.5in]{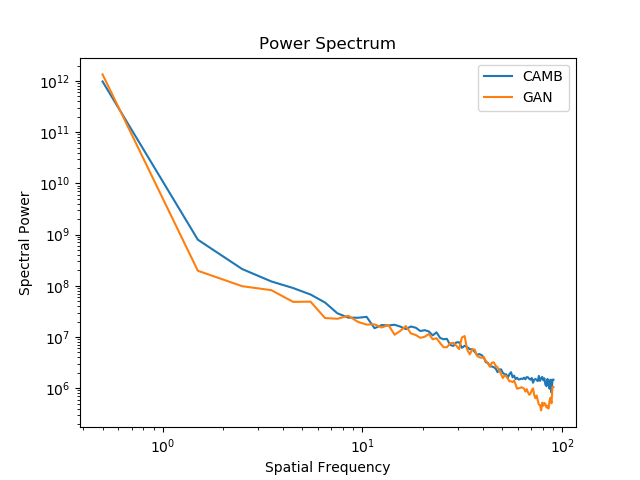}
	\caption{The power spectrum of the 2D image of a random sample patch from a map generated by CAMB and of a patch generated by trained GAN model.}		 	
\end{figure}

\begin{figure}[!h]
	\centering
	\includegraphics[width = 3.5in]{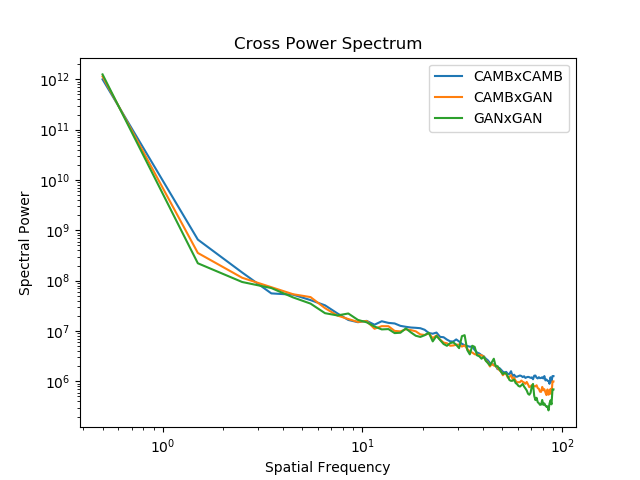}
	\caption{Cross Power Spectrum obtained using pairs of CAMB patches, pairs of GAN patches and between a CAMB and a GAN patch.}		 	
\end{figure}

\begin{figure}[!h]
	\centering
	\includegraphics[width = 3.5in]{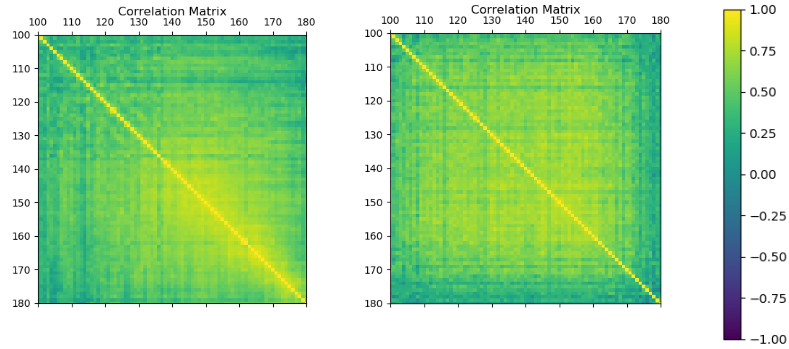}
	\caption{Correlation Matrix of the power spectrum of 100 GAN samples (right) and the corresponding matched CAMB samples (left).}		 	
\end{figure}

\newpage
\begin{figure}[!h]
	\centering
	\includegraphics[width = 3.5in]{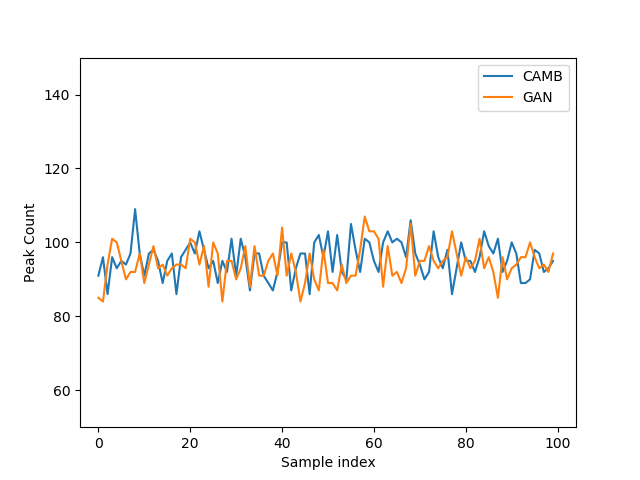}
	\caption{Peak Count of patches generated using GAN and the corresponding matched CAMB patches.}		 	
\end{figure}

\begin{figure}[!h]
	\centering
	\includegraphics[width = 3in]{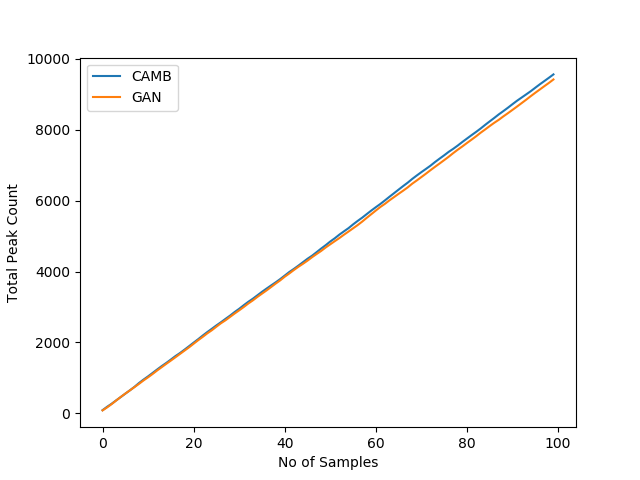}
	\caption{Total peak count of patches generated using GAN and the corresponding matched CAMB patches.}		 	
\end{figure}

\newpage
\subsection{112$^{\circ}$ patches}

\begin{figure}[!h]
	\centering
	\includegraphics[width = 3in]{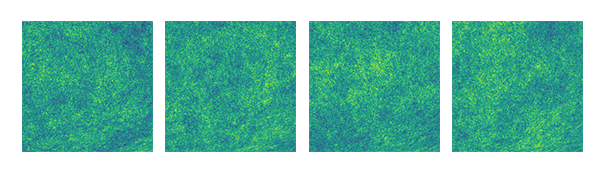}
	\caption{Sample patches (112$^{\circ}$s) from the maps generated using CAMB.}		 	
\end{figure}

\begin{figure}[!h]
	\centering
	\includegraphics[width = 3in]{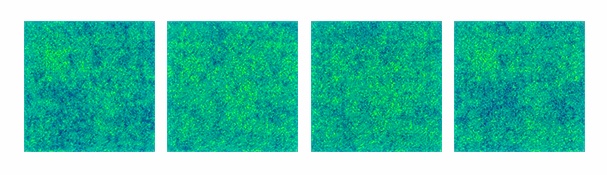}
	\caption{Sample patches (112$^{\circ}$s) generated by trained GAN model.}		 	
\end{figure}

\begin{figure}[!h]
	\centering
	\includegraphics[width = 3in]{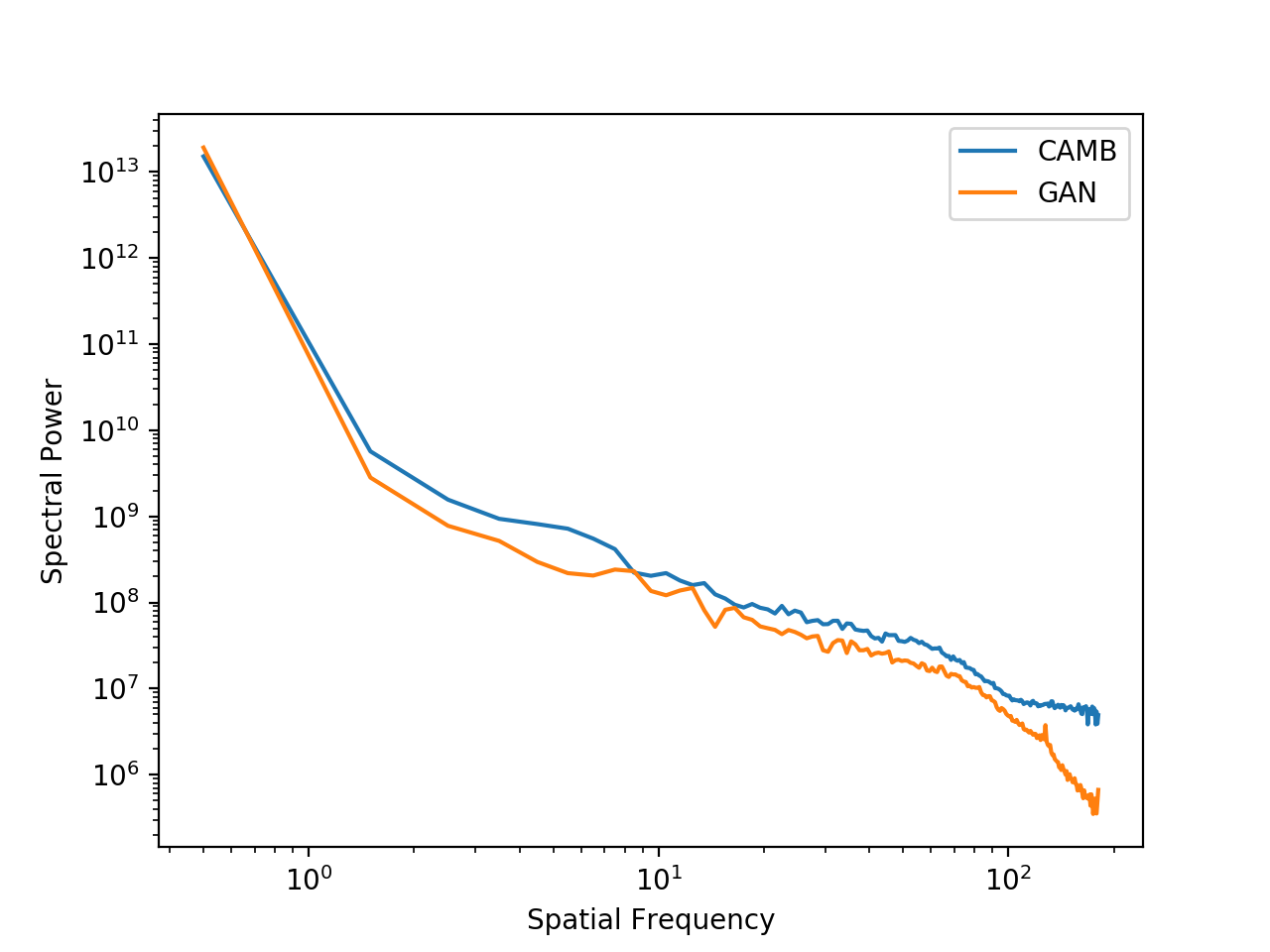}
	\caption{The power spectrum of the 2D image of a random sample patch from a map generated by CAMB and of a patch generated by trained GAN model.}		 	
\end{figure}

\begin{figure}[!h]
	\centering
	\includegraphics[width = 3in]{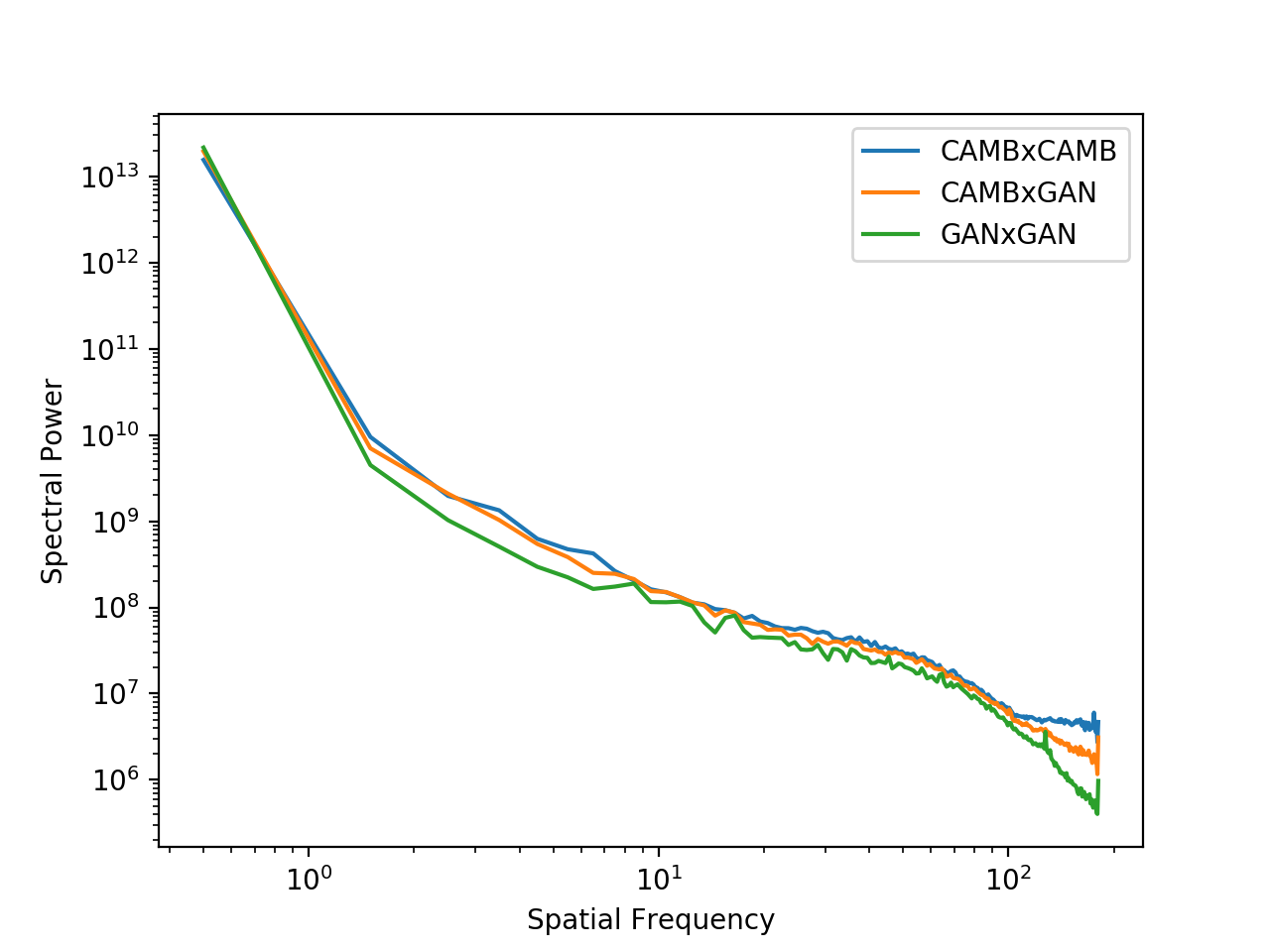}
	\caption{Cross Power Spectrum obtained using pairs of CAMB patches, pairs of GAN patches and between a CAMB and a GAN patch.}		 	
\end{figure}

\begin{figure}[!h]
	\centering
	\includegraphics[width = 3in]{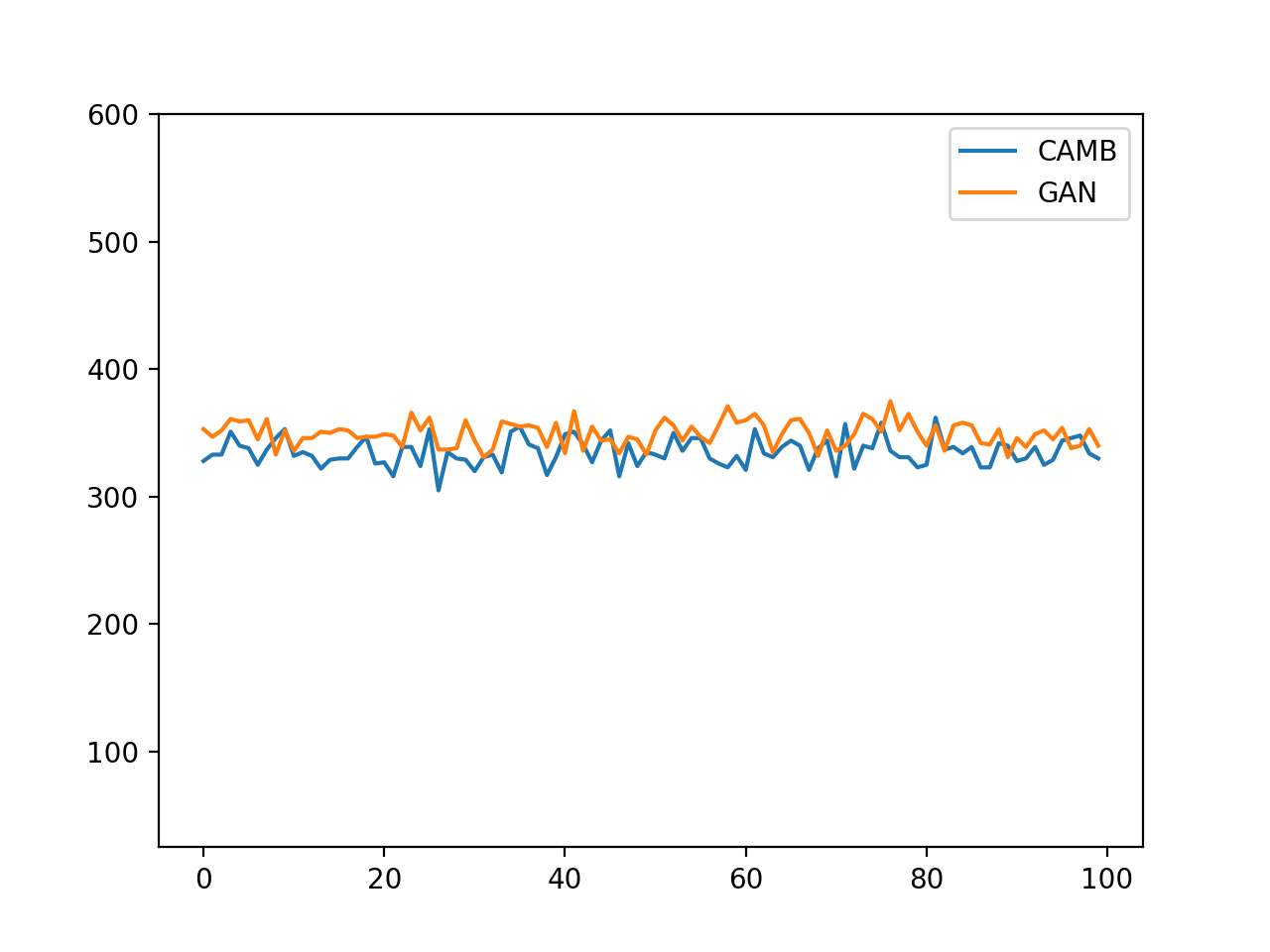}
	\caption{Peak Count of patches generated using GAN and the corresponding matched CAMB patches.}		 	
\end{figure}

\begin{figure}[!h]
	\centering
	\includegraphics[width = 3in]{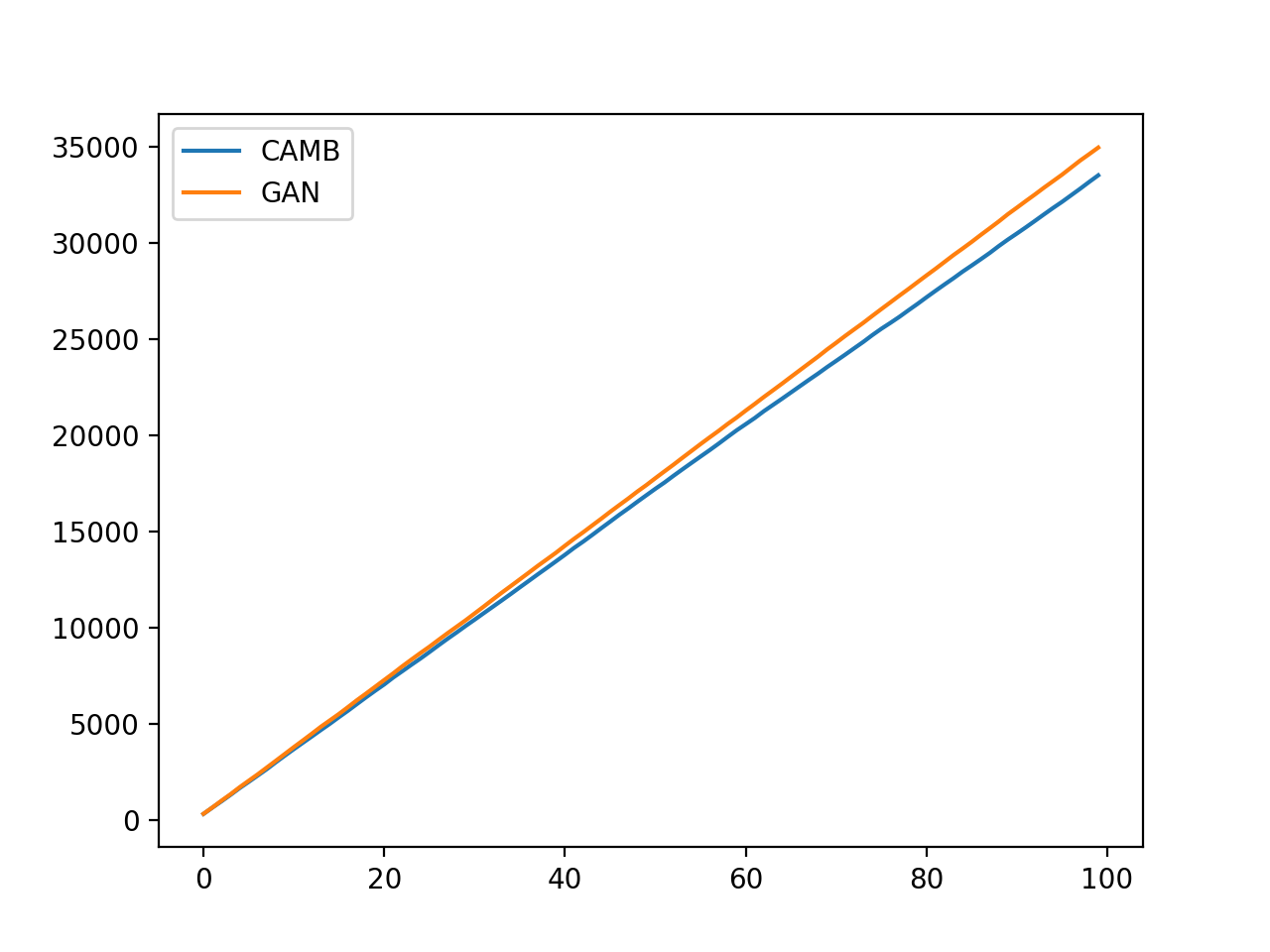}
	\caption{Total peak count of patches generated using GAN and the corresponding matched CAMB patches.}		 	
\end{figure}

\newpage
The random samples shown in Fig 13 and the subsequent results shown in Fig 14-17 follow a similar trend as we have seen with the results obtained from the model trained on 56◦ patches. This shows that the model is invariant to scale and is able to recreate the distribution of larger structures and can probably be extended to the generation of full-sky CMB maps. The higher deviation in the diagnostic metrics compared to the results of 56◦ patches can be attributed to the fact that we have used 2D projections of spherical maps and we can observe a loss of spatial information when we take larger projections. This can be solved by training the model with spherical CMB patches instead of 2D images and replacing the convolutional layers in the model with spherical convolutional layers.  

\section{Parameter variation}
To check the invariance of our model with respect to the change in the parameters corresponding to the input data that is used for training the models, we have trained two GAN models on CAMB data with Hubble constants 65 and 75 respectively. We have then used the trained model to generate 100 random samples for comparison. As seen in Fig 18, the standard deviation of the pixel intensity distributions of the samples with H = 65 are consistently higher than the ones obtained from the samples with H = 75, the average standard deviation of ( H = 65 ) samples is 15.6312 whereas that of the ( H = 75 ) samples is 14.2705 giving us a relative difference of 9.16\%.

Relative difference : \[\frac{|H_1-H_2|}{[\frac{(H_1+H_2)}{2}]}\times 100\]

We can also observe a noticeable variation in the correlation matrix (Fig 19) of samples of both the models. This shows that the model is sensitive to the change in the cosmological parameters that define the underlying distribution.

\begin{figure}[!h]
	\centering
	\includegraphics[width = 3in]{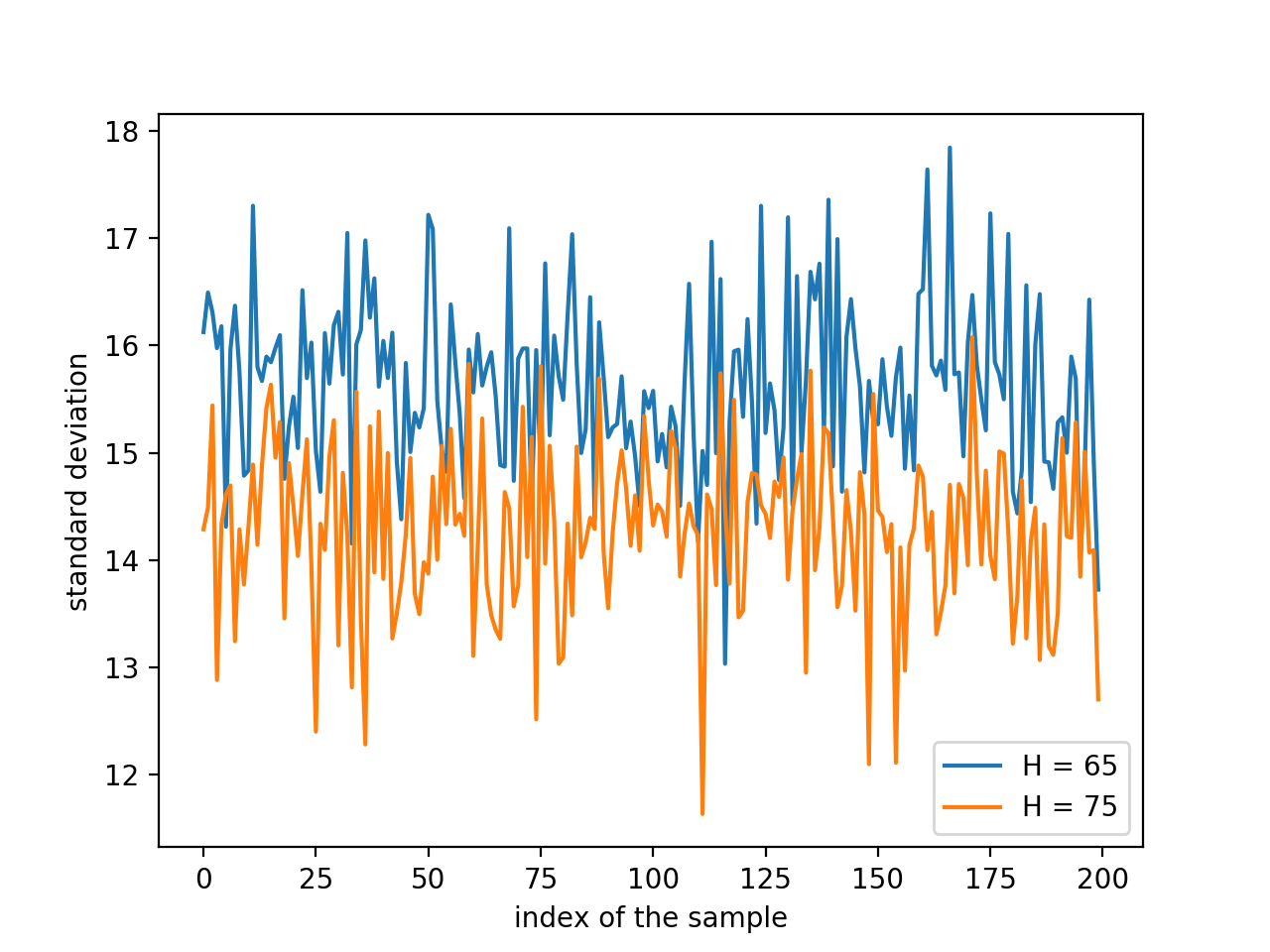}
	\caption{The  plot  of  the  standard  deviation  of  the  pixel intensity distribution of patches generated using GAN model trained on H = 65 data and H = 75 data.}
\end{figure}

\begin{figure}[!h]
	\centering
	\includegraphics[width = 3.5in]{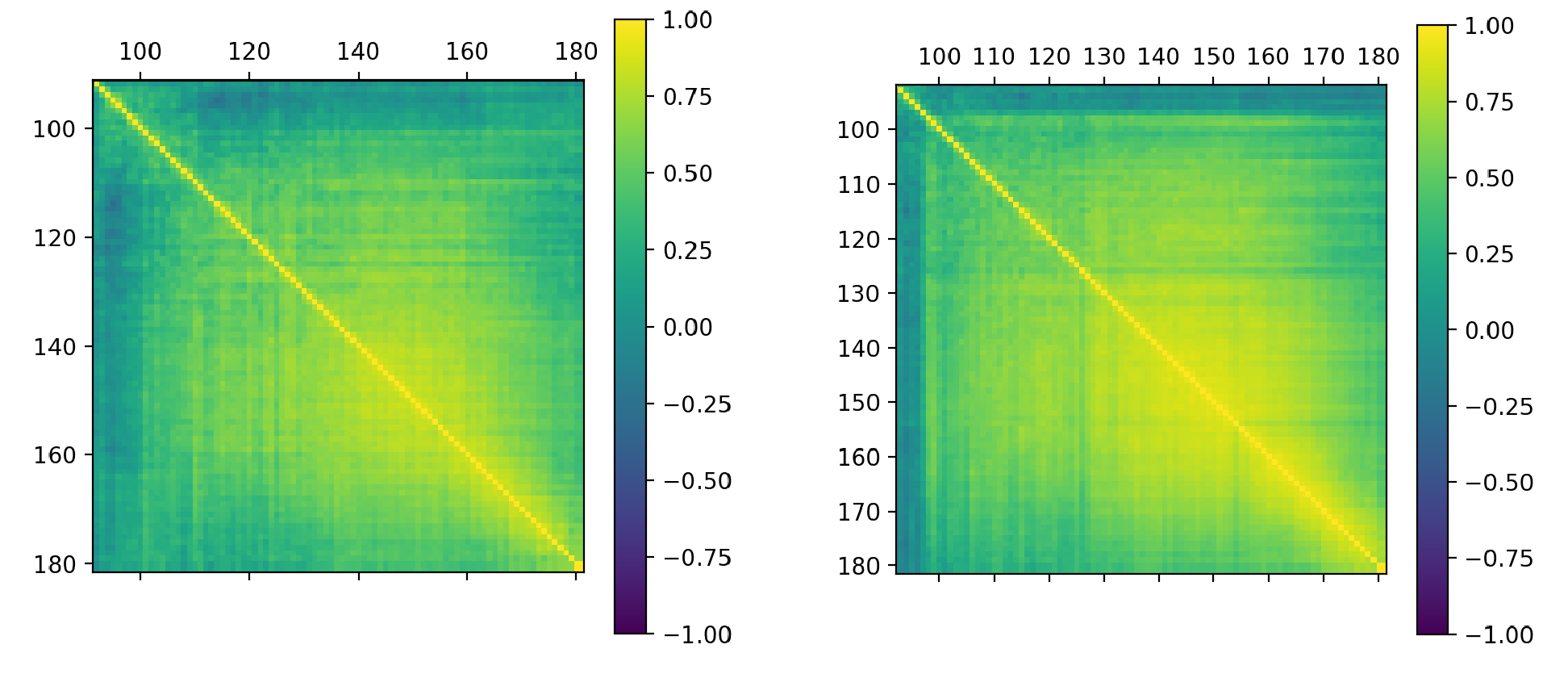}
	\caption{Correlation matrix of power spectrum of  GAN samples obtained from H = 65 model on the left and GAN samples obtained from H = 75 model on the right}
\end{figure}

We have also trained our model using 6 different values of Hubble's constant. Apart from an anomaly of H=73, as seen in Table 4 and Fig 20, the standard deviation of pixel intensity distribution appears to be decreasing with increase in Hubble's constant.

\begin{table}[!h]
\small
\caption{Results}
\begin{tabular}{| >{\centering\arraybackslash}m{1in} | >{\centering\arraybackslash}m{2.2in} |}
\hline
Hubble Constant & Average standard Deviation in pixel intensity distribution
\\
\hline
65  & 15.63120985989387\\
67  & 15.602640108842536\\
69  & 15.226652267061267\\
71  & 15.008699629159153\\
73  & 15.093389253196806\\
75  & 14.27058023246263\\
\hline
\end{tabular}
\end{table}

\begin{figure}[!h]
	\centering
	\includegraphics[width = 3in]{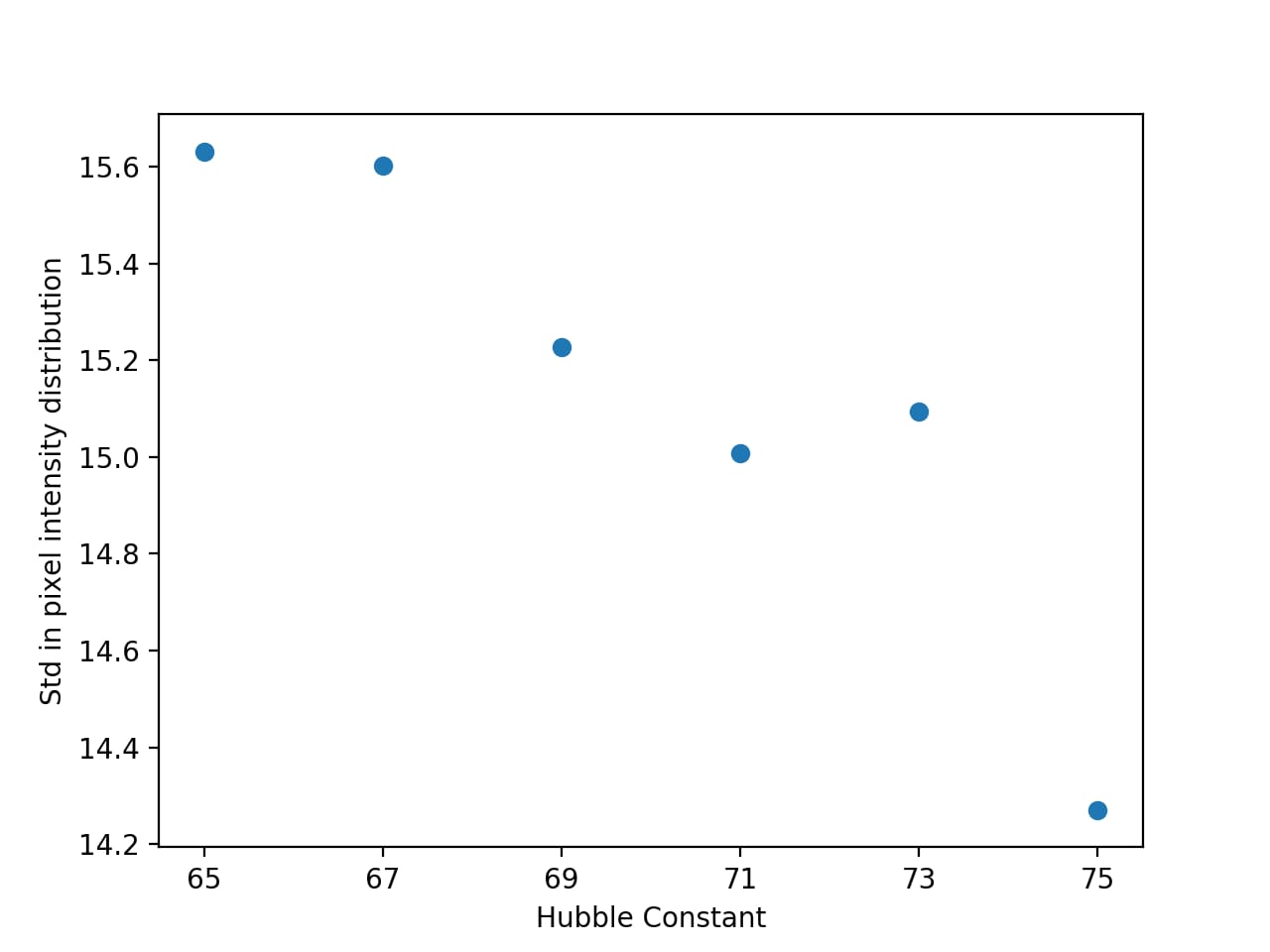}
	\caption{The  plot  of  the  average standard  deviation of the  pixel intensity distribution of 200 random samples vs the Hubble constant value corresponding to the training data}
\end{figure}

\newpage

As seen in Table 5 and Fig 21, we have also found the values of average standard deviation in the uncertainty of CMB Temperature using the scaling factor found from the average value of the mean of pixel intensity (luminosity) values and the average value of the mean of CMB temperature uncertainty values obtained from the temperature maps.

\begin{table}[!h]
\small
\caption{Results}
\begin{tabular}{| >{\centering\arraybackslash}m{1in} | >{\centering\arraybackslash}m{2.2in} |}
\hline
Hubble Constant & Average standard Deviation in uncertainty of CMB Temperature (mK)
\\
\hline
65  & 0.0544641458533\\
67  & 0.0543645996824\\
69  & 0.0530545375159\\
71  & 0.0522951206591\\
73  & 0.0525902064571\\
75  & 0.0497232760713\\
\hline
\end{tabular}
\end{table}

\begin{figure}[!h]
	\centering
	\includegraphics[width = 3in]{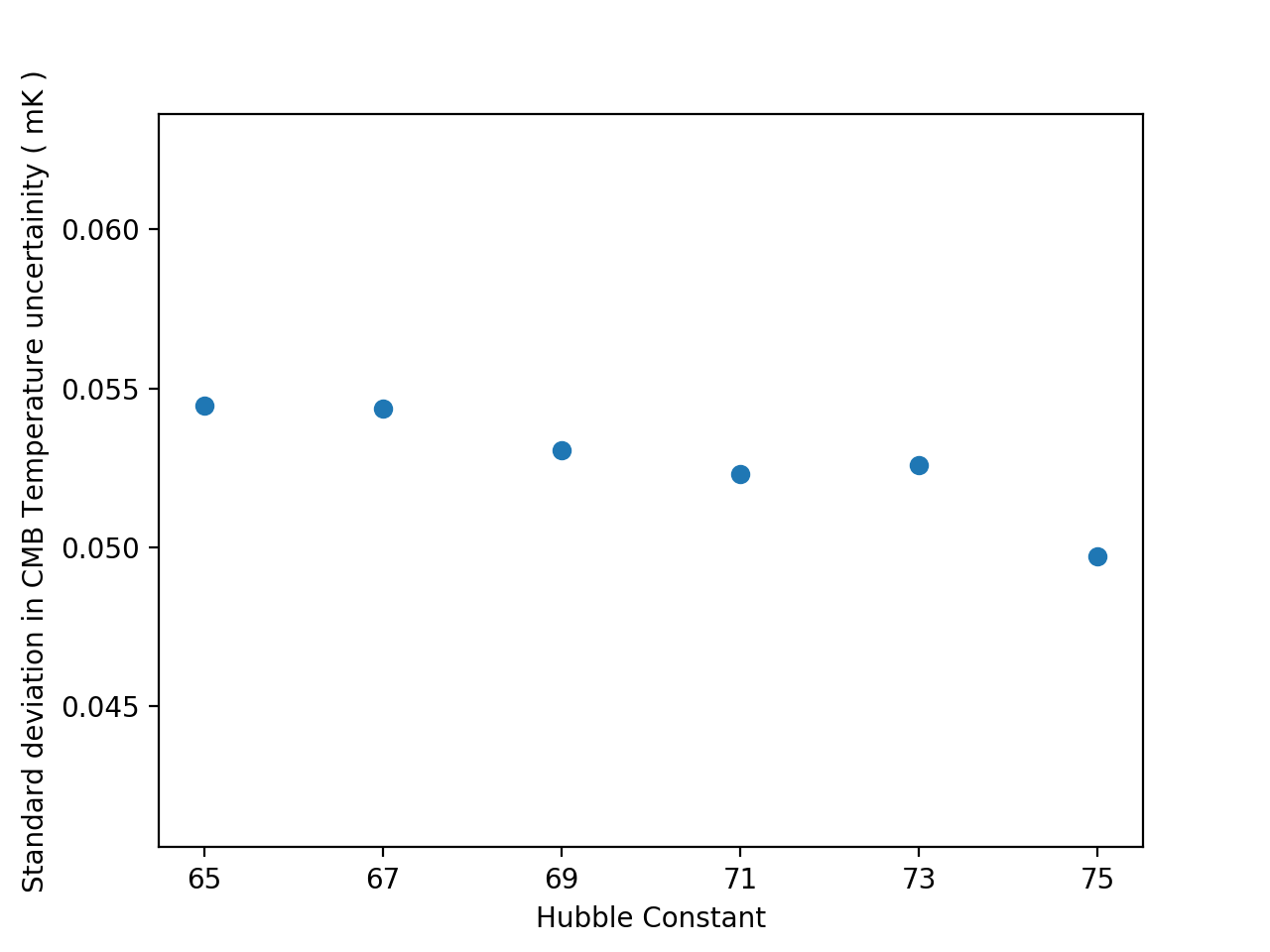}
	\caption{The  plot  of  the  average standard  deviation of the uncertainty of CMB Temperature of 200 random samples vs the Hubble constant value corresponding to the training data}
\end{figure}

We see a decreasing trend in standard deviation of uncertainty in CMB temperature with increasing Hubble's constant. This is expected since a higher Hubble's constant means a younger universe which in turn means the universe had lesser number of structures formed and result in lesser dispersion of relativistic particles and hence the lesser standard deviation in uncertainty in CMB temperature.

\section{Conclusion and future plans}
We have successfully presented the ability of Generative adversarial networks to learn the complex distribution behind flat CMB anisotropy maps. We have trained a deep convolutional generative adversarial network on a dataset of 56◦ CMB patches and 112◦ CMB patches obtained using CAMB. The patches generated by the GAN models are very similar to our training data, that are, the patches obtained by CAMB and healpy. The power spectrum of the patches generated by GAN and the patches obtained by CAMB are in very close agreement and a similar trend can be seen in other diagnostic metrics as well. We have also seen that our model is invariant to scale, that is, the size of the patch that has been used for training. The models trained on data with different values of Hubble constant have generated patches with significantly different properties, showing that our model is sensitive to change in cosmological parameters. We have shown that deep learning can be a viable alternative to traditional methods of CMB data generation and computationally much more efficient for cosmological experiments that require a large amount of CMB data, we hope to extend this study to the simulation of full-sky maps using spherical convolutional layers and generative adversarial networks in the future. We have trained our deep generative model using patches of CMB instead of full-sky maps because we were constrained by limited computational power. We hope to further improve our model and train using full-sky maps to present a viable method of CMB data generation for cosmological analysis.

\end{document}